# Joint Channel Assignment and Opportunistic Routing for Maximizing Throughput in Cognitive Radio Networks

Yang Qin*, Xiaoxiong Zhong*, Yuanyuan Yang+, Yanlin Li* and Li Li*

*Key Laboratory of Network Oriented Intelligent Computation, Shenzhen Graduate School,
Harbin Institute of Technology, Shenzhen, 518055, P. R. China

+Department of Electrical and Computer Engineering, Stony Brook University, Stony Brook, NY 11794, USA

Email: {xixzhong, yqinsg, liyanlin0}@gmail.com, yuanyuan.yang@stonybrook.edu, lili8503@aliyun.com

*Abstract*-In this paper, we consider the joint opportunistic routing and channel assignment problem in multi-channel multi-radio (MCMR) cognitive radio networks (CRNs) for improving aggregate throughput of the secondary users. We first present the nonlinear programming optimization model for this joint problem, taking into account the feature of CRNs-channel uncertainty. Then considering the queue state of a node, we propose a new scheme to select proper forwarding candidates for opportunistic routing. Furthermore, a new algorithm for calculating the forwarding probability of any packet at a node is proposed, which is used to calculate how many packets a forwarder should send, so that the duplicate transmission can be reduced compared with MAC-independent opportunistic routing & encoding (MORE) [11]. Our numerical results show that the proposed scheme performs significantly better that traditional routing and opportunistic routing in which channel assignment strategy is employed.

## I. INTRODUCTION

The cognitive radio principle has introduced the idea to exploit spectrum holes (i.e., bands) which result from the proven underutilization of the electromagnetic spectrum by modern wireless communication and broadcasting technologies [1]. CRNs have emerged as a prominent solution to improve the efficiency of spectrum usage and network capacity. In CRNs, secondary users (SUs) can exploit channels when the primary users (PUs) currently do not occupy the channels. The set of available channels for SUs is instable, varying over time and locations, which mainly depends on the PU's behavior. Thus, it is difficult to create and maintain the multi-hop paths among SUs through determining both the relay nodes and the available channels to be used on each link of the paths.

Taking the advantage of the broadcast nature and special diversity of the wireless medium, a new routing paradigm, known as opportunistic routing (OR) [2], has been proposed in the ExOR protocol. Instead of first determining the next hop and then sending the packet to it, a node with OR broadcasts the packet so that all neighbors of the node have the chance to hear it and assist in forwarding. OR provides significant throughput gains compared to traditional routing. In CRNs, it is hard to maintain a routing table due to dynamic spectrum access. The pre-determined end-to-end routing does not suit for CRNs either. Since opportunistic routing does not need prior setup of the route, it is more suitable for CRNs with dynamic changes of channel availability depending on the PU's behavior.

The effects of opportunistic routing on the performance of CRNs have been investigated in [3-8]. In 2008, Pan *et al.* [3] proposed a novel cost criterion for OR in CRNs, which leverages the unlicensed CR links to prioritize the candidate nodes and optimally selects the forwarder. In this scheme, the network layer selects multiple next-hop SUs and the link layer chooses one of them to be the actual next hop. The candidate next hops are prioritized based on their respective links' packet delivery rate, which in turn is affected by the PU activities. At the same time, Khalife *et al.* [4] introduced a novel probabilistic metric towards selecting the best path to the destination in terms of the spectrum/channel availability capacity. Considering the spectrum availability time, Badarneh *et al.* [5] gave a novel routing metric that jointly considers the spectrum availability of idle channels and the required CR transmission times over those channels. This metric aims at maximizing the probability of success (PoS) for a given CR transmission, which consequently improves network throughput. Lin *et al.* [6] proposed a spectrum aware opportunistic routing for single-channel CRNs that mainly considers the fading characteristics of highly dynamic wireless channels. The routing metric takes into account transmission, queuing and link-access delay for a given packet size in order to provide guarantee for end-to-end throughput requirement. Taking heterogeneous channel occupancy patterns into account, Liu *et al.* [7] introduced opportunistic routing into the CRNs where the statistical channel usage and the physical capacity in the wireless channels are exploited in the routing decision. Liu *et al.* [8] further discussed how to extend OR in multi-channel CRNs based a new routing metric, referred to as Cognitive Transport Throughput (CTT), which could capture the potential relay gain of each relay candidate. The locally calculated CTT values of the links (based on the local channel usage statistics) are the basis for selecting the next hop relay with the highest forwarding gain in the Opportunistic Cognitive Routing (OCR) protocol over multi-hop CRNs.

However, none of the above schemes systematically combines the channel assignment with OR to model CRNs. The number of candidate forwarders and the performance of OR will decrease, if using existing channel assignment algorithms for MCMR OR. A Workload-Aware Channel Assignment algorithm (WACA) for OR is designed in [9].

WACA identifies the nodes with high workloads in a flow as bottlenecks, and tries to assign channels to these nodes with high priority. WACA is the first static channel assignment for OR. However, it deals with channel assignment for single flow. Assuming that the number of radios and the number of channels are equal, a simple channel assignment for opportunistic routing (SCAOR) is proposed in [10]. It selects a channel for each flow. SCAOR is for multiple flows but assumes that the number of radios and the number of channels are equal. Neither of them is a feasible solution for OR in MCMR CRNs due to channel uncertainty of SUs. In this paper, we combine channel assignment and opportunistic routing, and analyze the impact of PU's behavior and buffer size on throughput.

The contributions of this paper can be summarized as follows. First, we propose a new scheme to select proper forwarding candidates for opportunistic routing, which considers the queue state of a node and channel availability. Second, a new algorithm for calculating the forwarding probability of any packet at a node is proposed, which is used to calculate the number of packets a forwarder should send. Finally, we formulate an optimization problem for combining opportunistic routing and channel assignment for CRNs, and compare the performance of our scheme, intra-session network coding-based opportunistic routing (ORNC), with shortest path routing (SINGLE), MORE [11], and ExOR in CRNs under different number of channels and buffer size.

The rest of this paper is organized as follows. In Section II, we describe the CRNs model used in this paper. In Section III, we formulate an optimization problem for joint opportunistic routing and channel assignment for CRNs. The numerical results are presented in Section IV. Finally, Section V concludes this paper.

## II. SYSTEM MODEL

We model the CRNs as a directed graph denoted by $G=(V,E)$, where $V$ is the set of $N$ SUs and $E$ is the set of links connecting any pair of nodes. The source node is denoted as $S$, and the destination is denoted as $D$. We consider a time slotted CRNs with $K$ licensed orthogonal channels belonging to an interweave model [12]. There are $N$ SUs and $M$ PUs in this CRN. Each node is equipped with the same number of radios $R$ in half-duplex model. Each SU is capable of sensing the locally available channels and has the capability of channel changing at packet level for data transmission. In CRNs, the SU's transmission range is $d_s$ and the interference range is $d_I$. Let $d_{ij}$ denote the distance between node $i$ and node $j$. If $d_{ij} < d_s$, we say nodes $i$ and $j$ are neighbors. Node $i$ and node $j$ can communicate with each other if they are neighbors and they are operating on the same channel. In OR, each node $i$ has multiple candidate forwarders denoted as $CFS_i$. For any two nodes, $i$ and $j$, $i<j$ indicates that node $i$ is closer to the destination node than node $j$, or in other words, $i$ has a smaller *ETX* (expected transmission count) [13] than $j$.

We summarize the notations used in this paper in TABLE I.

TABLE I
SUMMARY OF KEY NOTATIONS

| Symbol | Meaning |
| --- | --- |
| $G=(V,E)$ | the CRN topology graph |
| $S$ | the source node |
| $D$ | the destination node |
| $\psi_i^{k+}$ | the set of node $i$'s in-edge on the channel $k$ |
| $\psi_i^{k-}$ | the set of node $i$'s out-edge on the channel $k$ |
| $\rho_{ij}^k$ | the loss rate of link $e_{ij}$ ($e_{ij} \in E$) on the channel $k$ |
| $O_i^k(t)$ | the probability that node $i$ can transmit data packets using channel $k$ on time slot $t$ |
| $\theta_i^k(t)$ | the probability that node $i$ can use the channel $k$ on time slot $t$ |
| $P_i^k$ | the amount of packets that node $i$ has sent on channel $k$ |
| $\mu_{ij}^k(t)$ | the probability of $e_{ij}$ that transmits data packets using the channel $k$ on time slot $t$ |
| $f_{ij}^k$ | the number of data packets that $e_{ij}$ transmits on channel $k$ |
| $B$ | the maximum transmission rate on a channel |

## III. PROBLEM FORMULATION

In this section, we formulate the problem of joint channel assignment and OR as a nonlinear programming problem.

Let $h_{ij}^k \in \{0,1\}$ denote whether node $i$ and node $j$ can communicate with each other through channel $k$. If $h_{ij}^k = 1$, it means that nodes $i$ and $j$ can communicate with each other and $h_{ij}^k = 0$, vice versa.

We adopt the protocol interference model [14]. If $d_{uj} \leq d_s$, it means that node $j$ is in $u$'s transmission range. When nodes $i$ and $u$ simultaneously transmit data packets, the transmission of $i$ to $j$ will interfere with the transmission of $u$ to $v$ in time slot $t$. Similarly, when $d_{iv} \leq d_s$, the transmission of node $i$ to node $j$ will interfere with the transmission of nodes $u$ to $v$ in time slot $t$. Thus, we can calculate the interference link set $I_{ij}$ of link $e_{ij}$, $I_{ij} = \{<u,v> | <u,v> \in E, d_{uj} \leq d_s \text{ or } d_{iv} \leq d_s\}$, and have

$$\mu_{ij}^k(t) + \mu_{uv}^k(t) \leq 1, <u,v> \in I_{ij} \quad (1)$$

where $u_{ij}^k(t)$ is the probability of link $e_{ij}$ that transmits data packets using the channel $k$ in time slot $t$. In CRNs, $u_{ij}^k(t)$ is affected by the PU's activity. If two links are concurrently usable at the same channel in time slot $t$, they should either share the same transmitter or not interfere with each other. Hence, we can obtain

$$\sum_{<m,n> \in I_{ij}} \mu_{mn}^k(t) \leq 1, <m,n> \in I_{ij}, <i,j> \in E \quad (2)$$

Channel $k$ can be allocated to link $e_{ij}$ in time slot $t$ only when channel $k$ is available. Thus, we have

$$\mu_{ij}^k(t) \leq h_{ij}^k, <i,j> \in E \quad (3)$$

For each node $i$, it can participate in at most $R$ simultaneous communications in any given time $T$ ($T$ includes some mini-slots $t$). This can be formally represented by

$$\begin{cases} \mu_{ij}^k(t) \leq \theta_i^k(t), <i,j> \in \psi_i^{k-} \\ \mu_{gi}^k(t) \leq \theta_i^k(t), <g,i> \in \psi_i^{k+} \\ \sum_k \theta_i^k(t) \leq R \\ 0 \leq \theta_i^k(t) \leq 1 \end{cases} \quad (4)$$

where $\theta_i^k(t)$ is the probability that node $i$ can use the channel $k$ in time slot $t$, $\psi_i^{k+}$ is the set of node $i$'s in-edge on channel $k$, and $\psi_i^{k-}$ is the set of node $i$'s out-edges on channel $k$.

In MORE, the candidate forwarder is selected according to *ETX*. However, since in real world the buffer of a node is limited, it is reasonable to consider the buffer size in packet forwarding scheme. Thus, we should take buffer size constraint into account to select forwarding candidate. During time slot $T$, node $i$ sends $P_i^k$ packets on channel $k$. Then the queue length of $i$ at time $(T+1)$, $Q_i(T+1)$, can be expressed as

$$Q_i(T+1) = Q_i(T) - \sum_k P_i^k + (\sum_k \sum_t \sum_{<g,i> \in \psi_i^{k+}} f_{gi}^k \times (1-\rho_{gi}^k) \times \mu_{gi}^k(t)) \quad (5)$$

where $\rho_{gi}^k$ is the loss rate of link $e_{gi}$ on channel $k$, $f_{gi}^k$ is the number of data packets that $e_{gi}$ transmits on channel $k$, $P_i^k$ is the number of packets that node $i$ has sent on channel $k$ during time slot $T+1$, and $(\sum_k \sum_t \sum_{<g,i> \in \psi_i^{k+}} f_{gi}^k \times (1-\rho_{gi}^k) \times \mu_{gi}^k(t))$ is the amount of receiving packets of node $i$ during time slot $T+1$. Considering the queue backlog of node $i$, we use the summation of queue backlog and *ETX* as the forwarding candidate selector criterion, which can be expressed as

$$H_i(T) = \chi Q_i(T) + \gamma ETX_i \quad (6)$$

where $\chi$ and $\gamma$ are the weights, constrained by $\chi + \gamma = 1$, which are set to be 0.5 and 0.5 in our analysis.

In (6), we consider the queue length and channel availability in forwarding scheme. The smaller $H_i(T)$ node $i$ has, the higher probability the node to be selected as a forwarding candidate. If the queue of a node is almost full, which means packet loss will occur at the node, the node should not receive more packets.

For a given time $T$, the incoming packets of node $i$ are the same as the outgoing packets of node $i$ on channel $k$ to keep traffic balance. Also, the total number of data packets that $e_{ji}$ transmits on channel $k$ are not exceeding the maximum transmission rate of the channel. Therefore, we have

$$\begin{cases} 0 \leq f_{ij}^k \leq \sum_t \mu_{ij}^k(t) \times B, <i,j> \in E \\ \sum_k \sum_{<g,i> \in \psi_i^{k+}} f_{gi}^k \times (1-\rho_{gi}^k) \times \alpha_{gi}^k = \sum_k P_i^k + Q_i(T), i,g \in V, i \neq S \end{cases} \quad (7)$$

where $B$ is the maximum transmission rate on a channel, and $\alpha_{ij}^k$ is the forwarding probability that node $j$ forwards the packet received from node $i$ over channel $k$, to $j$'s next hop. Generally, the channel availability is heterogeneous in CRNs due to PU's activity. So, in our scheme, in each intermediate node, we attach the forwarding probability $\alpha_{ij}^k$ to each data packet, which can reduce duplicate transmission.

In the following, we give the algorithm to calculate $\alpha_{im}^k$.

**Algorithm 1** Calculate $\alpha_{im}^k$ in $CFS_i^k$

1: $\beta_{im}^k \leftarrow 0$, $temp \leftarrow 0$, $A \leftarrow \varnothing$
2: **for all** node $m$ in $CFS_i^k$ **do**
3:     calculate the probability $\beta_{im}^k(1)$ according to (8)
4:     calculate the probability $\beta_{im}^k(2)$ according to (9)
5:     $\beta_{im}^k \leftarrow \beta_{im}^k(1) + \beta_{im}^k(2)$
6:     $temp \leftarrow \beta_{im}^k + temp$
7: **end for**
8: **for all** node $m$ in $CFS_i^k$ **do**
9:     $\alpha_{im}^k \leftarrow \beta_{im}^k / temp$
10:     $A \leftarrow A \cup \{\alpha_{im}^k\}$
11: **end for**
12: **return** $A$

The probability that only node $m$ has received the packet is

$$\beta_{im}^k(1) = \rho_{i1}^k \rho_{i2}^k \rho_{im-1}^k (1-\rho_{im}^k) \rho_{im+1}^k \cdots \rho_{il}^k \quad (8)$$

where $\rho_{i1}^k$ corresponds to node 1 whose $ETX_1$ is the least in $CFS_i^k$, $\rho_{i2}^k$ is node 2 whose $ETX_2$ is in the second place, and so on, and $l$ is the number of candidates in $CFS_i^k$. All nodes, from 1 to $l$, are ordered by their *ETX*.

The probability that node $m$ and at least one node in $CFS_i^k$ ($m+1,\ldots l$) have received the packet is

$$\beta_{im}^k(2) = \rho_{i1}^k \rho_{i2}^k \rho_{im-1}^k (1-\rho_{im}^k)(1-\rho_{im+1}^k \rho_{im+2}^k \cdots \rho_{il}^k) \quad (9)$$

We now give an example for the algorithm. In Fig. 1, there are four nodes, 1, 2, 3 and 4, in node 0's $CFS_0^k$ on channel $k$ in a time slot, and the nodes are ordered by their *ETX*. The packet loss rate of link (0, 1) is $\rho_{01}$, similarly, $\rho_{02}$ for link (0, 2), $\rho_{03}$ for link (0, 3), and $\rho_{04}$ for link (0, 4).

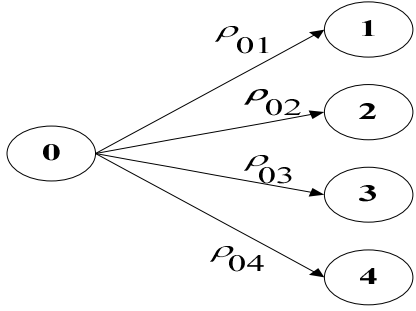

Fig. 1. An example for calculating $\alpha_{ij}^k$.

According to **Algorithm 1**, we can obtain

$$\begin{cases} \beta_{01}^k = (1-\rho_{01})\rho_{02}\rho_{03}\rho_{04} + (1-\rho_{01})(1-\rho_{02}\rho_{03}\rho_{04}) \\ \beta_{02}^k = \rho_{01}(1-\rho_{02})\rho_{03}\rho_{04} + \rho_{01}(1-\rho_{02})(1-\rho_{03}\rho_{04}) \\ \beta_{03}^k = \rho_{01}\rho_{02}(1-\rho_{03})\rho_{04} + \rho_{01}\rho_{02}(1-\rho_{03})(1-\rho_{04}) \\ \beta_{04}^k = \rho_{01}\rho_{02}\rho_{03}(1-\rho_{04}) \end{cases} \quad (10)$$

Node 1 will forward the packet received from node 0 with probability $\beta_{01}^k / \sum_{i=1}^{4} \beta_{0i}^k$ to its next hop. Similarly, node 2 will forward the packet with $\beta_{02}^k / \sum_{i=1}^{4} \beta_{0i}^k$, node 3 will forward the packet with $\beta_{03}^k / \sum_{i=1}^{4} \beta_{0i}^k$, and node 4 will forward the packet with $\beta_{04}^k / \sum_{i=1}^{4} \beta_{0i}^k$.

Note that our scheme adopts network coding [15, 16, 17, 18, 19, 20], and the coding operations are similar to the intra-session network coding in MORE. And, similar to WACA, we maintain $|K|$ credit counters for each node. Each credit counter corresponds to a channel. In our scheme, we consider the impact of channel availability of CRNs on credit calculating, which is shown as

$$credit_i^k = \frac{P_i^k}{\sum_t (\sum_{<g,i>\in \psi_i^{k+}} f_{gi}^k \times (1-\rho_{gi}^k) \times \mu_{gi}^k(t))}, i, g \in V, i \neq S \quad (11)$$

In Eq. (11), the parameter $\mu_{gi}^k(t)$ is the channel availability depending on PU's behavior in CRNs. If the $credit_i^k$ becomes positive, the node creates a coded packet, broadcasts it on channel $k$, and then decrements the credit counter.

In this paper, our goal is to maximize the aggregate throughput at destination node $D$. Thus, putting all the above constraints together, the objective function of the formulation is expressed as

$$\max \sum_k \sum_{<j,D>\in \psi_D^{k+}} f_{jD}^k \times (1-\rho_{jD}^k) \quad (12)$$

s.t.

$$\begin{cases} 0 \leq u_{ij}^k(t) \leq 1, <i,j>\in E \\ \sum_{<m,n>\in I_{ij}} \mu_{mn}^k(t) \leq 1, <m,n>\in I_{ij}, <i,j>\in E \\ u_{ij}^k(t) \leq \theta_i^k(t), <i,j>\in \psi_i^{k-}, i,j \in V, i \neq D \\ u_{gi}^k(t) \leq \theta_i^k(t), <g,i>\in \psi_i^{k+}, g,i \in V, i \neq S \\ \sum_k \theta_i^k(t) \leq R \\ 0 \leq \theta_i^k(t) \leq 1 \\ 0 \leq \alpha_{gi}^k \leq 1 \\ 0 < \rho_{ij}^k < 1 \\ u_{ij}^k(t) \leq h_{ij}^k, <i,j>\in E \\ 0 \leq f_{ij}^k \leq \sum_t u_{ij}^k(t) \times B, <i,j>\in E \\ \sum_k \sum_{<g,i>\in \psi_i^{k+}} f_{gi}^k \times (1-\rho_{gi}^k) \times \alpha_{gi}^k = \sum_k P_i^k + Q_i(T), g,i \in V, i \neq S \\ h_{ij}^k \in \{0,1\} \\ \forall k \in K, t \in T, i,j,g \in V \end{cases} \quad (13)$$

This is a nonlinear programming problem, and we can use the IBM ILOG CPLEX 12. 2 [21] to solve it.

## IV. NUMERICAL RESULTS

We compare the aggregate throughput among ORNC, SINGLE, ExOR and MORE in CRNs under different number of channels and buffer size. In the simulation, we randomly deploy 30 SUs and 4 PUs in a rectangle area of 500 units by 500 units. The interference range of SUs, $d_I$, is 8 units. The transmission range of SUs, $d_s$, is 4 units, while interference range of PUs is 12 units and the transmission range is 6 units. Each $T$ includes 5 time slots. The batch size is 10. Each link capacity is set to be 100 units. We set the packet loss rate $\rho_{ij}^k \in [0.1, 0.3]$.

In Fig. 2 and Fig. 3, the buffer size is 100 units. We can see that as the number of channels increases, there would be more available channels for CRNs, so the throughput of each scheme increases. ORNC, ExOR and MORE achieve higher throughput than SINGLE. The reason is that these three schemes take advantage of the inherent property of OR-opportunistic forwarding by using multiple forwarding candidates, while SINGLE always uses the same route consisting of a forward candidate. ORNC performs better than ExOR and MORE. This is because we exploit a new method for selecting forwarding candidates for ORNC in CRNs, which considers the queue length and channel availability. In addition, network coding in ORNC can reduce the retransmissions over forwarders' data transmission. The X3R (X: ORNC, MORE, ExOR, SINGLE) schemes, as shown in Fig. 2, perform clearly better than X1R, as shown in Fig. 3 which exploits possible concurrent transmissions by multi-radio nodes over the orthogonal channels of CRNs.

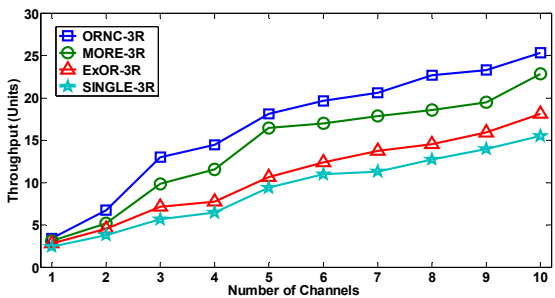

Fig. 2. Throughput comparison vs. Number of channels-with 3 radios.

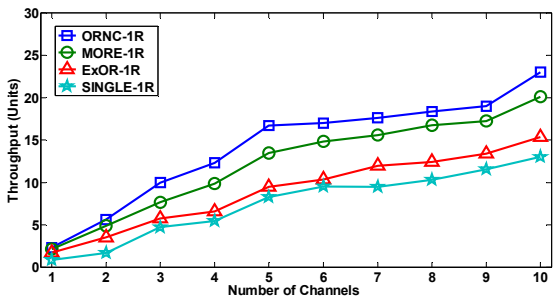

Fig. 3. Throughput comparison vs. Number of channels-with 1 radio.

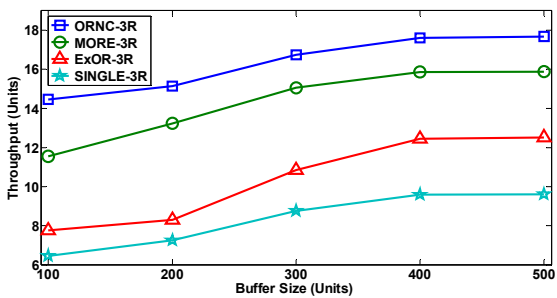

Fig. 4. Throughput comparison vs. Buffer size.

In Fig. 4 ($R=3$, $K=4$), we can see that as buffer size increases, the throughput increases and finally the growth is slowed down. This is due to the fact that for large buffer size, the packet loss is low. However, when the buffer size goes up to a certain value, the throughput increases slowly, as shown in Fig. 4. It is observed that the ORNC achieves much higher throughput than MORE, ExOR and SINGLE. This is because it considers the queue length and channel availability for selecting forwarding candidates in ORNC; also it exploits network coding technology in ORNC, which can reduce the retransmissions over multi-hop CRNs.

## V. CONCLUSION

In this paper, a novel scheme, ORNC, that jointly considers channel assignment and OR, is proposed for maximizing the aggregate throughput in multi-hop CRNs. In ORNC, we present a novel routing metric by considering queue state, *ETX* of a node and channel uncertainty. In addition, we propose a new algorithm for calculating the forwarding probability of any packet at a node that a packet can be sent at the node, which can reduce the duplicate transmission compared with MORE. And then, we formulate the joint problem as a nonlinear programming problem, and use ILOG CPLEX Optimizer to solve it. It is validated by numerical results that the proposed joint scheme ORNC achieves higher throughput than SINGLE, ExOR and MORE considering channel assignment. In future work, we will consider how to deal with the congestion in CRNs under opportunistic routing scenario.

## ACKNOWLEDGMENT

This work was supported by the Science and Technology Fundament Research Fund of Shenzhen under grant JC200903120189A,JC201005260183A, ZYA201106070013A. We would like to acknowledge the reviewers whose comments and suggestions significantly improved this paper.